\documentclass[preprintnumbers,article,amsmath,amssymb,floatfix,10pt,prd,onecolumn,
superscriptaddress,nofootinbib]{revtex4}
\usepackage{bbm}
\usepackage{amsfonts}
\usepackage{mathrsfs}
\usepackage{latexsym}

\usepackage{epsfig}
\usepackage{epstopdf}
\usepackage{placeins}
\usepackage{float}
\usepackage{graphicx}
\usepackage{amssymb}
\usepackage{amsmath}
\usepackage{dcolumn}
\usepackage{bm}
\usepackage{color}
\usepackage{comment}
\usepackage{xcolor}

\begin{document}

\title{\bf Stellar Structures in $f(\mathcal{G})$ Gravity Admitting Noether Symmetries}

\author{M. Farasat Shamir}
\email{farasat.shamir@nu.edu.pk}
\author{Tayyaba Naz}
\email{tayyaba.naz@nu.edu.pk}\affiliation{National University of Computer and
Emerging Sciences,\\ Lahore Campus, Pakistan.}

\begin{abstract}
This work aims to investigate some possible emergence of relativistic compact stellar objects in modified $f(\mathcal{G})$ gravity using $\emph{Noether symmetry approach}$. For this purpose, we assume static spherically symmetric spacetime in the presence of isotropic matter distribution. We construct Noether symmetry generators along with associated conserved quantities by considering the standard choice of viable $f(\mathcal{G})$ gravity model i.e. $f(\mathcal{G})= \alpha\mathcal{G}^{n}$, where $\alpha$ is the model parameter. In particular, we use conservation relation acquired from the classical Noether approach by imposing some appropriate initial conditions to construct the metric potentials. The obtained  conserved quantity  play vital role in describing the stellar structure of compact stars. Moreover, by considering an appropriate numerical solution, some salient features of compact stellar structures like effective energy density, pressure, energy conditions, stability against equilibrium of the forces and speed of sound are discussed by assigning the suitable values of model parameter involved. Our study reveals that the compact objects in $f(\mathcal{G})$ gravity from $\emph{Noether symmetry approach}$ depend on the conserved quantity obtained and the model parameter $\alpha$. In nutshell, Noether symmetries are quite helpful to generate solutions that follow physically accepted phenomena. Moreover, we observed that these obtained solutions are consistent with the astrophysical observational data, which depicts the viability of our proposed Noether symmetric scheme.
\end{abstract}
\maketitle
\section{Introduction}

$\emph{Noether symmetry approach}$ is considered to be viable mathematical tool, which explores the exact solutions as well as evaluates the corresponding integral of motion known as conserved quantities relative to the symmetry generator.
This approach plays a pivotal role to reduce a nonlinear system of equations into a linear system of equations.
A lot of interesting work has been done by many researchers in recent decades, for more detail (see \cite{14}-\cite{20}).
Furthermore, this approach provides a proficient method to calculate the conserved quantities \cite{Noether}.
The conservation laws play significant role in examining the different physical phenomenon.
These conservation laws are the specific cases of the renowned Noether's theorem, which depicts the relation between the
symmetries and the conserved quantities. Noether's theorem demonstrates that ``every differentiable symmetry
of a physical system under the action corresponds to some conservation law".
In particular, this theorem provides the information regarding the
 conservation laws in theory of general relativity ($\mathcal{GR}$).
In addition to this, conservation laws of linear and angular momentum can be complectly described by the rotational and translational symmetries with the help of Noether's theorem \cite{Hanc}.

Among different gravitational theories, the theory which has attained eminence in the last few years is
modified Gauss-Bonnet gravity also known as $f(\mathcal{G})$ gravity \cite{Noj1}-\cite{Cog2}.
Here, $f(\mathcal{G})$ is a generic function of the Gauss-Bonnet invariant term. The shortcomings related to $f(\mathcal{R})$ theory of gravity in the frame of large expansion of universe were resolved by the additional factor of the Gauss-Bonnet invariant term.
The Gauss-Bonnet term has great importance,  as it supports in regularizing the gravitational action
and may be helpful to avoid the ghost contributions \cite{T.Chiba}.
Moreover, this modified theory is considered to be an extensive tool to reveal the enigmatic nature of universe, which is the
actual reason of late-time acceleration expansion of universe. In addition, this modified $f(\mathcal{G})$ gravity yields a valuable platform to investigate the behavior of future time singularities \cite{Santos, Sen}. For the study of late-time cosmic acceleration as well as the finite time future singularities, the modified $f(\mathcal{G)}$ theory is considered to be very helpful \cite{Noj2008, Bamba}. Uddin et al. \cite{eqiuv} studied the generalized
Gauss-Bonnet gravity theories that leads to such solutions where the universe is
governed by a barotropic perfect fluid. They argued that $f(\mathcal{G})$ theory of gravity may be interpreted as an effective `scalar-tensor'
theory. Further, it was shown by employing an equivalence between
the Gauss-Bonnet action and a scalar-tensor theory of gravity that the cosmological
field equations could be written as a plane autonomous system.

The $\emph{Noether symmetry approach}$ has been widely used in modified theories of gravity,
which yield some significant solutions for the cosmological systems.
The energy contents of stringy black hole solutions in the presence of charge were explored by
Sharif and Waheed \cite{Sharif1} with the help of $\emph{Noether symmetry approach}$.
Kucukakca \cite{Kucu} found the exact solutions of Bianchi type-$\textbf{I}$ model by using Noether symmetries method.
In \cite{Cap2007}, $\emph{Noether symmetry approach}$ was used to find the exact solution of Friedmann-Robertson-Walker (FRW) spacetime in the frame of $f(\mathcal{R})$ theory of gravity. Bahamonde et al. \cite{Bahamonde} considered $\emph{Noether symmetry approach}$, to find a class of new exact solutions of the spherically symmetric spacetime. Moreover by assuming the Noether symmetries method, the exact solutions of the field equations were constructed by Shamir and Ahmad \cite{Mush1, Mush2} in the context of $f(\mathcal{G,T})$ gravity.
$\emph{Noether symmetry approach}$ has been taken into account by Sharif and Fatima \cite{Sharif2} to examine the cosmic evolution as well as cosmic expansion. Laurentis \cite{Laurentis} adopted $\emph{Noether symmetry approach}$ to construct the spherically symmetric solutions in $f(\mathcal{R})$ gravity.
Bahamonde and his collaborators \cite{b} used Noether symmetries as a selection criterion to investigate some $f(\mathcal{R,G})$ gravity models that are invariant under point transformations in a spherically symmetric background. In total, they found ten different functions that represent symmetries and calculated their invariant quantities. Exact solutions in modified teleparallel gravity for the cases of spherically and cylindrically symmetric tetrads have been discussed using Noether symmetries of point-like Lagrangians defined in Jordan and Einstein frames \cite{s}.
In a recent paper \cite{f}, a class of modified $f(\mathcal{R})$ theories of gravity has been studied in detail using $\emph{Noether symmetry approach}$ and is concluded that different scenarios of cosmic evolution can be discussed using by Noether symmetries and one of the case indicates the chances for the existence of Big Rip singularity. Moreover, the scalar field involved in the modified gravity plays a vital role in the cosmic evolution and an accelerated expansion phase can be observed for some suitable choices of scalar modified gravity models.

The phenomenon of self-gravitation and its consequences have great impact in the field of novel astrophysical research and cosmological background.
In the result of this gravitational collapse, some new stellar remnants known as compact stars are produced. These compact stars are considered to be very dense as they possess large masses but volumetrically small radii.
In relativistic astrophysics, compact stars have drawn the attention of the researcher
due to their captivating characteristics and relativistic structures.
These compact stars can be well explained by $\mathcal{GR}$ as well as the modified theories of gravity \cite{Abbas1}-\cite{Cam}.
Abbas et al. \cite{Abbas4} found the equilibrium condition for compact stars, further they also examined their physical characteristics in the context of modified $f(\mathcal{G})$ gravity. The role of modified theories of gravity is very eminent while studying and examining the structure of compact stellar objects and matter at high densities \cite{Ast2}-\cite{Ast4}.

Interesting discussions in above paragraphs motivate us to construct Noether symmetry generators and associated conserved quantities of spherically symmetric spacetime by assuming the viable $f(\mathcal{G})$ gravity model i.e. $f(\mathcal{G})= \alpha\mathcal{G}^{n}$ \cite{Cog2}, in the context of isotropic matter distribution. To our best of knowledge, this is the first effort to use conservation
relation obtained from the classical Noether approach, by imposing some specific initial conditions to develop the
expression for the metric potentials in the context of $f(\mathcal{G})$ gravity to discuss compact stellar structures.
The layout of this paper is as follows:
In section $\textbf{II}$, we present the mathematical formulation of $f(\mathcal{G})$ gravity in the context of isotropic matter distributions.
Section $\textbf{III}$ is devoted to construct a symmetry reduced Lagrangian, further a brief review of $\emph{Noether symmetry approach}$ is also presented in the same section. Section $\textbf{IV}$, is based on finding the expression for the metric potentials by making use of conservational relation with some appropriate initial conditions. Section $\textbf{V}$ is devoted to scrutinize some physical attributes of compact stellar structure and also for checking viability of model via graphical analysis. Last section is based on the conclusive remarks.

\section{Some Basics of $f(\mathcal{G})$ Gravity}

We consider the most general action for modified $f(\mathcal{G})$ gravity defined as \cite{Noj1}
\begin{equation}\label{5.1}
  \mathcal{S} = \int d^4x \sqrt{-g} \Bigg[\frac{\mathcal{R}}{2\kappa^2}+f(\mathcal{G})+L_m\Bigg],
\end{equation}
where $L_m$ is the matter Lagrangian, $\mathcal{R}$ is the Ricci scalar, $\kappa^2 = {8\pi G}$ represents the
coupling constant term which we consider as unity henceforth and $f(\mathcal{G})$ being an arbitrary function of
Gauss-Bonnet invariant term expressed as
\begin{equation}\label{6}
\mathcal{G} = \mathcal{R}^2 - 4\mathcal{R}_{\mu\nu}\mathcal{R}^{\mu\nu} +\mathcal{R}_{\mu\nu\sigma\rho}\mathcal{R}^{\mu\nu\sigma\rho},
\end{equation}
here $\mathcal{R}_{\mu\nu}$ and $\mathcal{R}_{\mu\nu\rho\sigma}$ indicate the Ricci and Riemann tensors, respectively.
The following modified field equations are obtained by varying the action (\ref{5.1}) with respect to metric tensor $g_{\mu\nu}$
\begin{eqnarray}\nonumber
G_{\mu\nu}+ 8\big[\mathcal{R}_{\mu\rho\nu\sigma} + \mathcal{R}_{\rho\nu}g_{\sigma\mu} - \mathcal{R}_{\rho\sigma}g_{\nu\mu} - \mathcal{R}_{\mu\nu}g_{\sigma\rho} + \mathcal{R}_{\mu\sigma}g_{\nu\rho} + \frac{\mathcal{R}}{2}(g_{\mu\nu}g_{\sigma\rho}-g_{\mu\sigma}g_{\nu\rho})\big]\nabla^{\rho}\nabla^{\sigma}f_\mathcal{G}+(\mathcal{G}f_\mathcal{G}- f)g_{\mu\nu} =\kappa^2T_{\mu\nu}.\\\label{5.2}
\end{eqnarray}
In the above equation, subscript $\mathcal{G}$ in $f_\mathcal{G}$ represents the derivative with respect to $\mathcal{G}$.
Another alternative form of modified field equations (\ref{5.2}), familiar with $\mathcal{GR}$ can be written as
\begin{eqnarray}\label{5.4}
G_{\mu\nu} =\kappa^2T_{\mu\nu}^{eff},
\end{eqnarray}
the effective stress-energy tensor $T_{\mu\nu}^{eff}$ is given by
\begin{eqnarray}\label{5.5}
T_{\mu\nu}^{eff}=T_{\mu\nu}- \frac{8}{\kappa^2}\big[\mathcal{R}_{\mu\rho\nu\sigma} + \mathcal{R}_{\rho\nu}g_{\sigma\mu} -\mathcal{R}_{\rho\sigma}g_{\nu\mu} - \mathcal{R}_{\mu\nu}g_{\sigma\rho} +\mathcal{R}_{\mu\sigma}g_{\nu\rho} + \frac{\mathcal{R}}{2}(g_{\mu\nu}g_{\sigma\rho}-g_{\mu\sigma}g_{\nu\rho})\big]\nabla^{\rho}\nabla^{\sigma}f_\mathcal{G}
 -(\mathcal{G}f_\mathcal{G}- f)g_{\mu\nu}.
\end{eqnarray}
Further, to investigate and examine the nature of compact stars, we choose a static spherically symmetric spacetime \cite{Krori}.
\begin{equation}\label{5.6}
ds^{2}= e^{\nu(r)}dt^2-e^{\lambda(r)}dr^2-r^2(d\theta^2+\sin^2 \theta d\phi^2).
\end{equation}
The source of matter configuration assumed in this present study is isotropic in nature and can be
expressed by the following energy momentum tensor
\begin{equation}\label{5.3}
T_{\mu\nu}=(\rho+p)u_\mu u_\nu-pg_{\mu\nu},
\end{equation}
where $u^\mu$ and $u_\nu$ are the radial-four vectors and four velocity vectors,
respectively. Using equations (\ref{5.5}) and (\ref{5.6}), we obtain the following set of field equations for the isotropic stellar system as
\begin{eqnarray}\label{10}
\rho^{eff}&&=~~\rho-8e^{-2\lambda}(f_\mathcal{GGG}\mathcal{G}'^{2}+f_\mathcal{GG}\mathcal{G}'')(\frac{e^{\lambda}-1}{r^2})
+4e^{-2\lambda}\lambda' \mathcal{G}'f_\mathcal{GG}(\frac{e^{\lambda}-3}{r^2})-(\mathcal{G}f_\mathcal{G}-f),
\end{eqnarray}
\begin{eqnarray}\label{11}
p^{eff}&&=~~ ~p-4e^{-2\lambda}\nu'\mathcal{G}'f_\mathcal{GG}(\frac{e^{\lambda}-3}{r^2})+
(\mathcal{G}f_\mathcal{G}-f).\quad\quad\quad\quad\quad\\\nonumber
\end{eqnarray}
Here $\rho$, $p$ are usual energy density and pressure respectively.
The system of equations have four unknown functions namely, $\rho$, $p$, $\lambda$, $\nu$.
For the above equations of motion, the expressions for $\rho^{eff}$ and $p^{eff}$  are equal to the corresponding components of the Einstein tensor.
\\\\$\mathbf{\textit{\textbf{Theorem}:}}$
Given a solution of Eqs. (\ref{10}) and (\ref{11}), defined by the functions $W_1=\big\{\nu(r), \lambda(r), f(\mathcal{G})\big\}$, if we have a solution in $\mathcal{GR}$ defined by $W_2=\big\{\nu(r), \lambda(r)\big\}$, then all the physical attributes are identical for $W_1$
and $W_2$ since $T^{eff}_{\mu\nu}$ in (\ref{5.4}) plays the role of stress-energy tensor in $\mathcal{GR}$ \cite{M.V}.
\\\\The Eqs. (\ref{10}) and (\ref{11}) are very much intricate and non-linear because of the involved variable function $f(\mathcal{G})$.

\section{Point-Like Lagrangian in $f(\mathcal{G})$ Gravity and Noether Symmetry Approach}

The physical properties of a dynamical system can be analyzed by the construction of associated Lagrangian
which clearly illustrates the possible existence of symmetries and  energy content.
So, in this situation $\emph{Noether symmetry approach}$ yields fascinating way to develop the new
cosmological models and geometries in modified theories of gravity.
Noether's theorem states that every differentiable symmetry of the action of a physical system
corresponds to some conservation law. In this section, we construct the point-like Lagrangian for
the static spherically symmetric metric in the framework of modified $f(\mathcal{G})$ gravity. By using the $\emph{Noether symmetry approach}$,
we compute the determining equations of the system. The existence of this approach assures the uniqueness of the vector field in the associated tangent space. In this way, the vector field acts like symmetry generator which further yields the conserved quantities of the dynamical system.
We express canonical form of action (\ref{5.1}) in such manner that the degrees of freedom are reduced.
\begin{equation}\label{8}
\mathcal{S} = \int dr\mathcal {L}(q^{i},{q'^{i}})=\int dr\mathcal {L}(\nu,{\nu'},\lambda,{\lambda'},\mathcal{G},{\mathcal{G'}}).
\end{equation}
We consider the Lagrange multiplier technique by taking $\mathcal{G}$ as a
dynamical constraint. By selecting the appropriate Lagrange multiplier and integrating by parts, we eliminate the second order derivatives in such a way that Lagrangian becomes canonical. Thus the action (\ref{5.1}) takes the form
\begin{equation}\label{5.7}
\mathcal{S} = \int dr\sqrt{-g}\Big[f-\gamma(\mathcal{G}-\bar{\mathcal{G}})\Big],
\end{equation}
where
$\sqrt{-g}=e^{\frac{\nu+\lambda}{2}}r^2\sin \theta$ and $\mathcal{G}=\frac{2e^{-\lambda}}{r^2}(\nu'\lambda'+{\nu'}^{2}e^{-\lambda}-3\nu'\lambda'e^{-\lambda}
-2\nu''-{\nu'}^{2}+2\nu''e^{-\lambda}).$
The term $\gamma=f_\mathcal{G}$ is called Lagrange multiplier and is obtained by varying the above action with respect to $\mathcal{G}$. Hence we can re-write the action (\ref{5.7}) as follows
\begin{equation}\label{5.8}
\mathcal{S} = \int dr\sqrt{-g}\Big[f-f_\mathcal{G}\Big(\mathcal{G}-\frac{2e^{-\lambda}}{r^2}\big\{\nu'\lambda'+{\nu'}^{2}e^{-\lambda}-3\nu'\lambda'e^{-\lambda}
-2\nu''-{\nu'}^{2}+2\nu''e^{-\lambda}\big\}\Big)\Big].
\end{equation}
The expression for the Lagrangian density is given as
\begin{equation}\label{5.9}
\mathcal{L}(\nu,{\nu'},\lambda,{\lambda'},\mathcal{G},{\mathcal{G'}})=e^{\frac{\nu+\lambda}{2}}r^2(f-\mathcal{G}f_\mathcal{G})+4e^{\frac{\nu-\lambda}{2}} \nu'\mathcal{G}'f_\mathcal{GG}-
4e^{\frac{\nu-3\lambda}{2}} \nu'\mathcal{G}'f_\mathcal{GG}.
\end{equation}
Further, the Euler-Lagrange equations are represented by
\begin{equation}\label{5.10}
\frac{\partial\mathcal{L}}{\partial q^i}-\frac{d}{dr}(\frac{\partial\mathcal{L}}{\partial {q'^{i}}})=0.
\end{equation}
For static spherically symmetric spacetime (\ref{5.6}) and (\ref{5.9}), the Euler-Lagrange equations turn out to be
\begin{eqnarray}\nonumber
&&2e^{\frac{\nu-\lambda}{2}}(\nu'-\lambda')\mathcal{G}'f_\mathcal{GG}
-2e^{\frac{\nu-3\lambda}{2}}(\nu'-3\lambda')\mathcal{G}'f_\mathcal{GG}
+4e^{\frac{\nu-\lambda}{2}}\mathcal{G}''f_\mathcal{GG}-4e^{\frac{\nu-3\lambda}{2}}\mathcal{G}''f_\mathcal{GG}
+4e^{\frac{\nu-\lambda}{2}}\mathcal{G}'^{2}f_\mathcal{GGG}\\\label{5.10}
&&-4e^{\frac{\nu-3\lambda}{2}}\mathcal{G}'^{2}f_\mathcal{GGG}
-2e^{\frac{\nu-\lambda}{2}}\nu'\mathcal{G}'f_\mathcal{GG}+2e^{\frac{\nu-3\lambda}{2}}\nu'\mathcal{G}'f_\mathcal{GG}
-\frac{1}{2}e^{\frac{\nu+\lambda}{2}}r^2(f-\mathcal{G}f_\mathcal{G})=0,
\end{eqnarray}
\begin{eqnarray}\label{5.11}
6e^{\frac{\nu-3\lambda}{2}}\nu'\mathcal{G}'f_\mathcal{GG}
-2e^{\frac{\nu-\lambda}{2}}\nu'\mathcal{G}'f_\mathcal{GG}+\frac{1}{2}e^{\frac{\nu+\lambda}{2}}r^2(f-\mathcal{G}f_\mathcal{G})=0,
\end{eqnarray}
\begin{eqnarray}\label{5.12}
&&2e^{\frac{\nu-\lambda}{2}}(\nu'-\lambda')\nu'f_\mathcal{GG}
-2e^{\frac{\nu-3\lambda}{2}}(\nu'-3\lambda')\nu'f_\mathcal{GG}
+4e^{\frac{\nu-\lambda}{2}}\nu''f_\mathcal{GG}-4e^{\frac{\nu-3\lambda}{2}}\nu''f_\mathcal{GG}
+e^{\frac{\nu+\lambda}{2}}r^2\mathcal{G}f_\mathcal{GG}=0.
\end{eqnarray}
The energy function $\mathcal{E}_{\mathcal{L}}=0$
associated with point-like Lagrangian is defined as
\begin{eqnarray}\label{5.13}
\mathcal{E}_{\mathcal{L}}=\Sigma {q'^{i}}(\frac{\partial\mathcal{L}}{\partial {q'^{i}}})-\mathcal{L}=0.
\end{eqnarray}
For our case the energy function associated with the Lagrangian (\ref{5.9}) leads to
\begin{eqnarray}\label{5.14}
\mathcal{E}_{\mathcal{L}}=4e^{\frac{\nu-\lambda}{2}}\nu'\mathcal{G}'f_\mathcal{GG}
-4e^{\frac{\nu-3\lambda}{2}}\nu'\mathcal{G}'f_\mathcal{GG}
-e^{\frac{\nu+\lambda}{2}}r^2(f-\mathcal{G}f_\mathcal{G}).
\end{eqnarray}
A generator of the form for which the point-like Lagrangian (\ref{5.9}) admits Noether symmetries is given by \cite{P.J}
\begin{eqnarray}\label{5.15}
\mathcal{Z}=\mathcal{\xi}\frac{\partial}{\partial r}+\eta^{i}\frac{\partial}{\partial q^{i}}.
\end{eqnarray}
Here the generalized coordinates $q^{i}$ in the $\emph{d}$-dimensional configuration space $\mathcal{Q}=\{q^{i},i=1,...$\emph{d}$\}$ of the Lagrangian, whose tangent space is $\mathcal{\tau}\mathcal{Q}\equiv \{q^{i},q'^{i}\}$.
It is important to mention here that the components $\mathcal{\xi}$ and $\eta^{i}$ of the Noether symmetry generator $\mathcal{Z}$ are the functions of $r$ and $q^{i}$.
In our case, $\mathcal{\xi}\equiv \mathcal{\xi}(\nu,\lambda,\mathcal{G})$ and $q^{i}\equiv q^{i}(r,\nu,\lambda,\mathcal{G})$, $i=\{1,...,3\}$. The existence of Noether symmetry is confirmed only if the Lagrangian $\mathcal{L}(r,q^{i},q'^{i})$ must satisfy invariance condition and the existence of vector field $\mathcal{Z}$ given in (\ref{5.15}) expressed in the following form
\begin{eqnarray}\label{5.16}
\mathcal{Z}^{[1]}\mathcal{L}+\mathcal{L}(D_{r} \mathcal{\xi})=D_{r}\mathcal{B},
\end{eqnarray}
 where $\mathcal{Z}^{[1]}$ is called the first prolongation of above generator (\ref{5.15}), given by
 \begin{eqnarray}\label{5.16}
 \mathcal{Z}^{[1]}= \mathcal{Z}+\eta'^{i}\frac{\partial}{\partial q'^{i}},
 \end{eqnarray}
and
\begin{itemize}
  \item $\mathcal{B}(r,q^{i})$ is known as Noether gauge term,
   \item  $D_r$ is a total derivative operator,
   $D_{r}=\frac{\partial}{\partial r}+q'^{i}\frac{\partial}{\partial q^{i}}$,
   \item $\eta'^{i}$ is defined as $\eta'^{i}=D_{r}\eta^{i}-q'^{i}D_{r}\mathcal{\xi}$.
\end{itemize}
The most significant feature known as integral of motion or the conserved quantity comes from the invariance property of Noether symmetry. The conserved quantity associated to any Noether symmetry generator $\mathcal{Z}$ is defined by
\begin{eqnarray}\label{5.16}
\mathcal{I}=-\mathcal{\xi}\mathcal{E}_{\mathcal{L}}+\eta^{i}\frac{\partial \mathcal{L}}{\partial q'^{i}}-\mathcal{B}.
\end{eqnarray}
\subsection{Determining Equations and Conserved Quantities}

The conserved quantities play significant role in constructing new cosmological models and describing the features of compact stars in modified theories of gravity.
Now by taking Lagrangian (\ref{5.9}) along Noether symmetry condition (\ref{5.16}), we obtain a set of partial differential equations (PDE's) known as determining equations by equating the coefficients. For our case we get a system of $11$ PDEs as follows
\begin{eqnarray}\label{5.17}
\xi,_{\nu}=0,~~~~~~~~~\xi,_{\lambda}=0,~~~~~~~\xi,_{\mathcal{G}}=0,~~~~~~~\eta^{1},_{\lambda}=0, ~~~~~~\eta^{1},_{r}=\mathcal{B},_{\mathcal{G}},
\end{eqnarray}
\begin{eqnarray}\label{5.18}
\eta^{1},_{\mathcal{G}}=0,~~~~~~~~~~~~~\eta^{3},_{r}=\mathcal{B},_{\nu},~~~~~~~~~~~~~~\eta^{3},_{\nu}=0,~~~~~~~~~~~~~\eta^{3},_{\lambda}=0,
\end{eqnarray}
\begin{eqnarray}\label{5.19}
\frac{1}{2}re^{\frac{\nu+\lambda}{2}}\Bigg[(r(\eta^{1}+\eta^{2})+4\xi+2r\xi,_{r})(f-\mathcal{G}f_\mathcal{G})-2\mathcal{G}r\eta^{3} f_\mathcal{GG}\Bigg]=\mathcal{B},_{r},
\end{eqnarray}
\begin{eqnarray}\label{5.20}
e^{\frac{\nu-3\lambda}{2}}\Bigg[2\Big\{2\xi,_r-\eta^{1}+3\eta^{2}+2(e^{\lambda}-1)\eta^{1},_\nu-2\eta^{3},_\mathcal{G}
+e^{\lambda}(\eta^{1}-\eta^{2}-2\xi,_r+2\eta^{3},_\mathcal{G})\Big\}f_\mathcal{GG}+4(e^{\lambda}-1)\gamma f_\mathcal{GG}\Bigg]=0.
\end{eqnarray}
Further, we solve the system of above PDEs for the proposed model.
Now we compute Noether symmetries $\mathcal{Z}=\mathcal{\xi}\frac{\partial}{\partial r}+\eta^{i}\frac{\partial}{\partial q^{i}}$ by solving the above system of PDEs. For this purpose, we consider the viable $f(\mathcal{G})$ gravity model i.e. $f(\mathcal{G})=\alpha \mathcal{G}^n$, where $f(\mathcal{G})$ being an analytic function of the Gauss-Bonnet invariant term \cite{Cog2}. Here we particulary choose $n=2$ for the sake of simplicity and two dimensional graphical analysis. Now manipulating Eqs. (\ref{5.17})-(\ref{5.20}), we get
\begin{eqnarray}\label{5.21}
\mathcal{\xi}=c_{1}r,~~~~~~~ \eta^{1}=6c_{1},~~~~~~~~\eta^{2}=0,~~~~~~
\eta^{3}=-4\mathcal{G}c_{1},~~~~~~~\mathcal{B}=c_{2}.
\end{eqnarray}
 The Noether symmetry generator turns out to be
 \begin{eqnarray}\label{5.21}
 \mathcal{Z}_{1}=r\frac{\partial}{\partial r}+6\frac{\partial}{\partial \nu}-4\mathcal{G}\frac{\partial}{\partial \mathcal{G}}.
\end{eqnarray}
Moreover, using Eq. (\ref{5.16}) the integral of motion for above symmetry generator becomes,
\begin{eqnarray}\label{5.22}
\mathcal{I}_{1}= e^{\frac{\nu-3\lambda}{2}}[-\mathcal{G}\{e^{2\lambda}\mathcal{G} r^3+32(e^{\lambda}-1)\nu'\}
-8(e^{\lambda}-1)(r\nu'-10)\mathcal{G}'].
\end{eqnarray}

\section{Noether's Star in $f(\mathcal{G})$ Gravity}

It seems very fascinating that the use of Noether symmetries is quite essential to ``reduce" dynamics by searching out the first integrals of motion known as conserved quantities. Moreover, these conserved quantities are linked with some physical parameters of dynamical system.
In nutshell, the resulting system is fully integrable, if the number of conserved quantities coincides with the dimension of the configuration space.
These quantities play an important role in exploring the different physical features of compact stars.  In particular, this technique is successfully applicable to dynamical systems in axial and spherical symmetry \cite{Cap2007, Cap2010}. The solutions at surface boundary of the compact stars
are calculated in such a way the interior metric solution should be smoothly matched with the corresponding exterior solution. The interior and exterior metric potentials are connected through the following relation
\begin{eqnarray}\label{5.23}
e^{\nu(r)} &=& e^{-\lambda(r)}.
\end{eqnarray}
For this purpose, we consider the above conserved quantity (\ref{5.22}) for finding the metric potentials i.e. $\nu$ and $\lambda$ which yields fruitful result in finding the physical attributes of the compact stars. Now using the relation (\ref{5.23}) in Eq. (\ref{5.22}), we get
\begin{eqnarray}\label{5.24}
\mathcal{I}_{1}= e^{2\nu}[-e^{2\nu}r^3\mathcal{G}^2 +32(e^{-\nu}-1)\nu'\mathcal{G}
-8(e^{-\nu}-1)(r\nu'-10)\mathcal{G}'].
\end{eqnarray}
One can see that the above Eq. (\ref{5.24}) is highly nonlinear differential equation and it does not support any information
to solve this equation analytically. Here we opt numerical technique depending on the computational efficiency and use some
appropriate initial conditions to investigate the physical behavior of metric potentials.
Now we solve the above Eq. (\ref{5.24}) numerically with suitable initial conditions.

\subsection{Boundary Conditions}

\begin{figure}[h!]
\begin{tabular}{cccc}
\epsfig{file=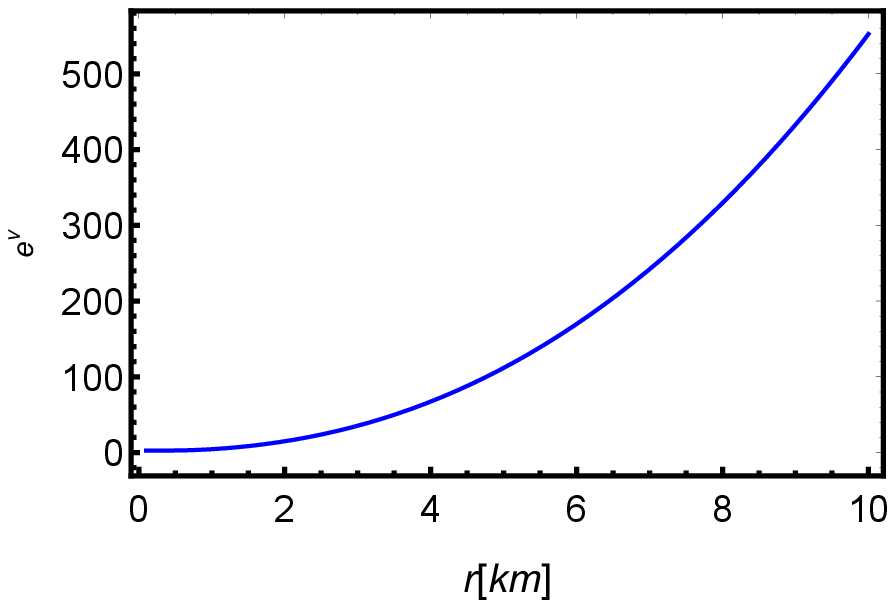,width=0.38\linewidth} &
\epsfig{file=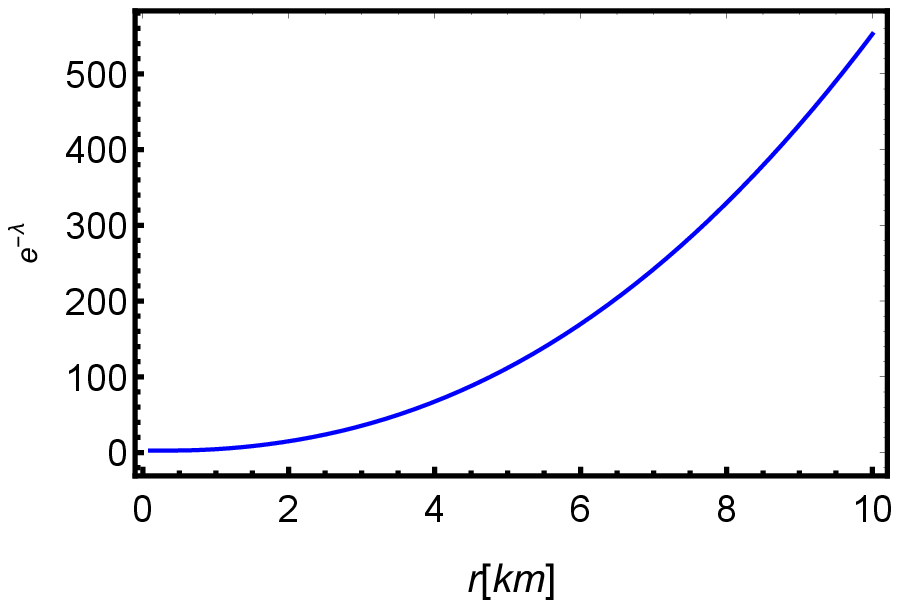,width=0.38\linewidth} &
\end{tabular}
\caption{The evolution of $e^{\nu}$ (left panel), $e^{\lambda}$ (right panel) w.r.t the radial coordinate $r$}.
\label{Fig:1}
\end{figure}

For the existence of singularities, we examine the physical behavior of both metric potentials  $e^{\nu(r)}$ and $e^{\lambda(r)}$
at the center of structure $r=0$.
The physical viability and stability of the model, metric potentials should be singularity-free, monotonically increasing and regular inside the compact stellar structure. The graphical behavior of metric potentials $e^{\nu}$ and $e^{\lambda}$ is shown in FIG. $\ref{Fig:1}$.
\FloatBarrier
It is very clear from the graphical analysis that the metric potentials constructed with the help of integral of motion are consistent with the above mentioned conditions. Both metric potentials at the center are minimum, then increase nonlinearly and become maximum at the boundary surface.
\\Moreover, it is worthwhile to mention here that for well behaved compact stellar objects, the following conditions should be satisfied in order to contrast the astrophysical observational data.
\begin{itemize}
\item The effective energy density must be positive inside the star and also on the surface boundary i.e. $\rho^{eff}>0$ and $0\leq r\leq R.$
\item  The isotropic pressure must be positive inside the star i.e. $p^{eff}>0$ and at the surface boundary $r=R$, the pressure must be zero i.e $p^{eff}(r=R)=0$.
    \item The effective energy density gradient $\frac{d^{eff}\rho}{dr}$ must be negative for the range $0\leq r\leq R$, i.e. $(\frac{d\rho^{eff}}{dr})_{r=0}=0$ and $(\frac{d^2\rho^{eff}}{dr^2})_{r=0}<0$.

 \item The effective pressure gradient $\frac{dp^{eff}}{dr}$ must be negative for the range $0\leq r\leq R$, i.e. $(\frac{dp^{eff}}{dr})_{r=0}=0$ and $(\frac{d^2p^{eff}}{dr^2})_{r=0}<0$.
These above two conditions depict that the effective energy density and effective radial pressure should be decreasing towards the boundary of the surface of the structure.

\item The velocity of sound speed must not be exceed the speed of light.

\end{itemize}
These physical features are significant in describing the structure of compact star.

\section{Physical Aspects in $f(\mathcal{G})$ Gravity}

The physical attributes of the compact stellar structure in modified $f(\mathcal{G})$ are
discussed in the following subsections. Here we present the physical analysis of the characteristics for the intrinsic region of the star by performing numerical calculations and graphical survey viz., effective energy density and effective pressure, energy conditions, stability against equilibrium of the forces and radial sound speed.

\subsection{Evolution of Energy Density and Pressure}

\begin{figure}[h!]
\begin{tabular}{cccc}
\epsfig{file=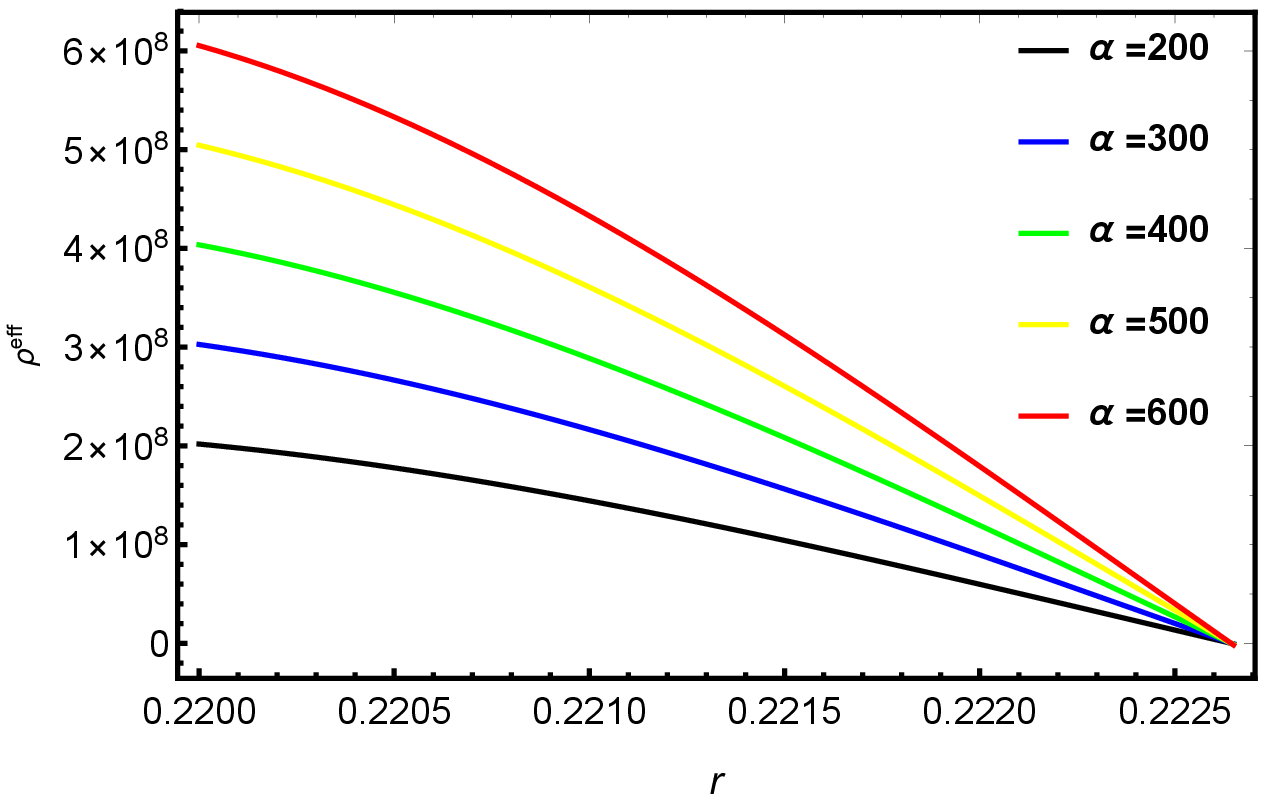,width=0.43\linewidth} &
\epsfig{file=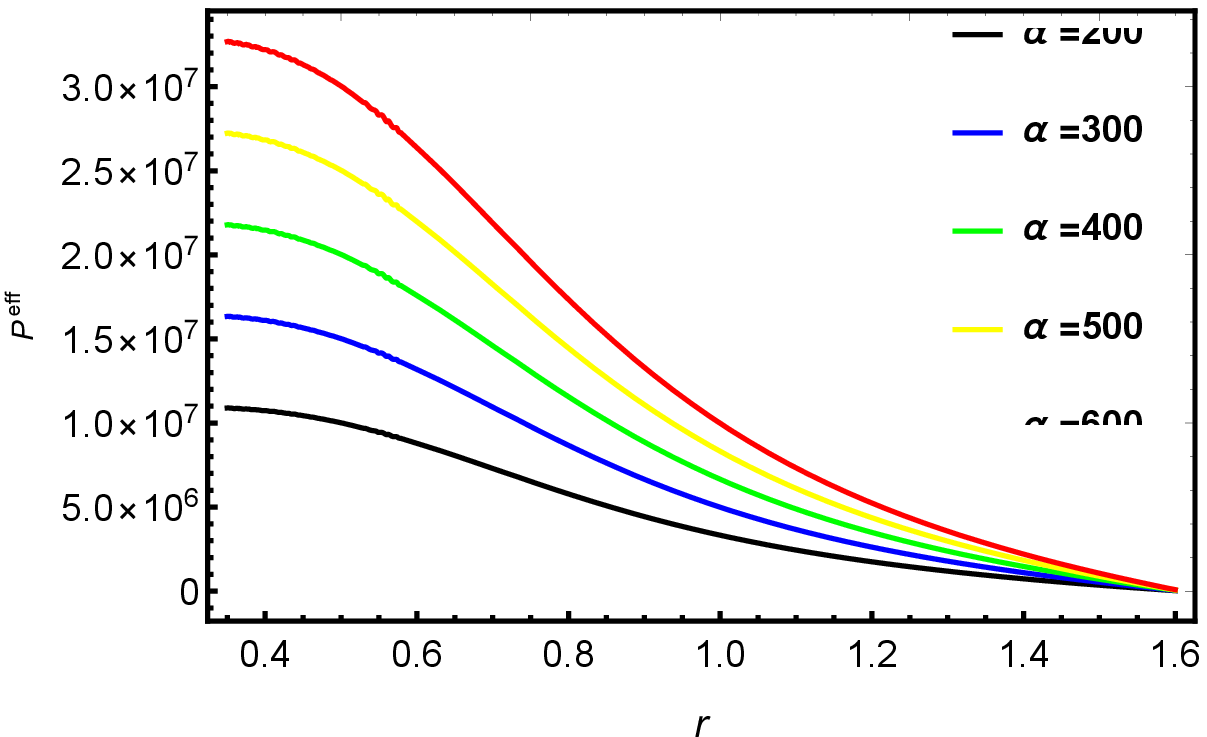,width=0.45\linewidth} &
\end{tabular}
\caption{Evolution of effective energy density (left panel), effective isotropic pressure (right panel) w.r.t the radial coordinate $r$.}
\label{Fig:2}
\end{figure}

The effective energy density and effective pressure
inside the stellar structure of star should be maximum at the center. To verify this, we had plotted the graphs for the range $0\leq r\leq 10$ to check the behavior of stellar structures. For this range, we get interweaving behavior for the physical features.
We have also tried log scale plot to improve the graphical behavior. But in this case also, the graphical results could not be improved considerably. In fact, we come to know that favorable results are obtained in small pieces of radii.
However, we observe that at center these features attain maximum value and then instantaneously decrease towards the surface boundary, which is a positive indication for physically viable results through Noether symmetries. Hence, for clarity we have provided only the plots for the small range of radii to show the smooth behavior of our stellar structures by using Noether scheme.
The profile of effective energy density and isotropic  pressure is shown in FIG. $\ref{Fig:2}$.
\FloatBarrier
It is observed from these graphs that effective energy density and isotropic  pressure are finite and  positive and show the monotonically decreasing behavior away from the center to the boundary of the star. This fact confirms the high compactness nature at the core of compact stellar object. Moreover, we notice that at the center of the star $\frac{d\rho^{eff}}{dr}=0$, $\frac{d^2\rho^{eff}}{dr^2}<0$ and $\frac{dp^{eff}_r}{dr}=0$, $\frac{d^2p^{eff}_{r}}{dr^2}<0$, which
depicts the dense nature of the star. Here we can observe that the stellar structures in $f(\mathcal{G})$ gravity from $\emph{Noether symmetry approach}$ heavily depend on the conserved quantity obtained.

\subsection{Energy Conditions}

\begin{figure}[h!]
\begin{tabular}{cccc}
\epsfig{file=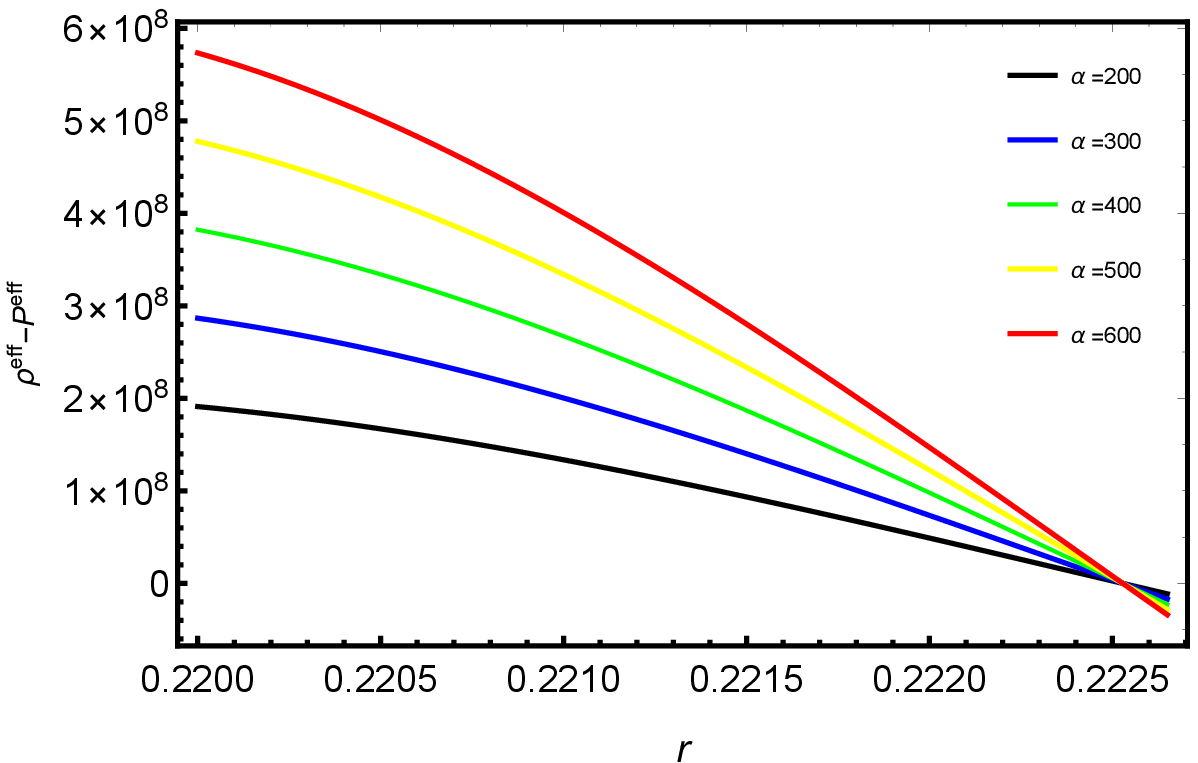,width=0.43\linewidth} &
\epsfig{file=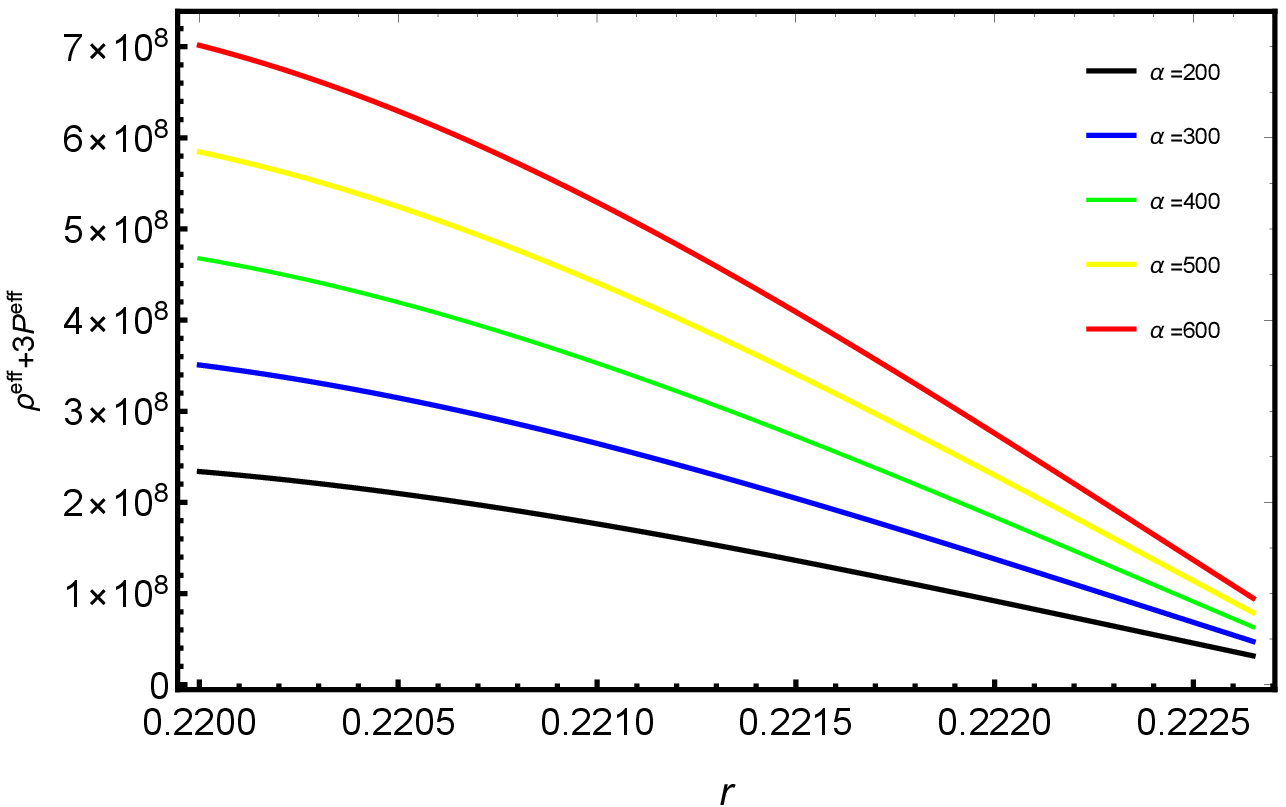,width=0.43\linewidth} &
\end{tabular}
\caption{Variation of energy conditions under viable $f(\mathcal{G})$ gravity model.}
\label{Fig:3}
\end{figure}

For the presence of realistic matter distribution and physically adequate fluid configuration, the role of energy conditions is very important.
Moreover, energy conditions are assumed quite helpful to identify the normal and exotic nature of matter inside the compact stellar structure.
These energy conditions are categorized into null energy conditions $\mathcal{(NEC)}$, weak energy conditions $\mathcal{(WEC)}$, strong energy conditions $\mathcal{(SEC)}$ and dominant energy conditions $\mathcal{(DEC)}$, expressed as
\begin{eqnarray}\nonumber
&&\mathcal{(NEC)}:~~~~~~\rho^{eff}+p^{eff}\geq0\nonumber,
\\&&\mathcal{(WEC)}:~~~~~~\rho^{eff}\geq0,~~~\rho^{eff}+p^{eff}\geq0,\nonumber,
\\&&\mathcal{(SEC)}:~~~~~~\rho^{eff}+p^{eff}\geq0,~~~\rho^{eff}+3p^{eff}\geq0,\nonumber
\\&&\mathcal{(DEC)}:~~~~~~\rho^{eff}>|p^{eff}|.
\end{eqnarray}
For the physical viability of stellar structure, the matter density should be
positive and finite throughout the star and also have maximum value at the
center. Here we also face the same interweaving behavior in the plots as discussed earlier. So, the graphical behavior of energy conditions for different values of model parameter is also given in FIG. $\ref{Fig:3}$ for small range of radii.
The graphical representation as shown in FIG. $\ref{Fig:3}$ clearly depicts that the above inequalities hold at every point inside the star, which shows that our chosen $f(\mathcal{G})$ model is stable and consistent.
\begin{figure}[h!]
\begin{tabular}{cccc}
\epsfig{file=densityN.eps,width=0.43\linewidth} &
\epsfig{file=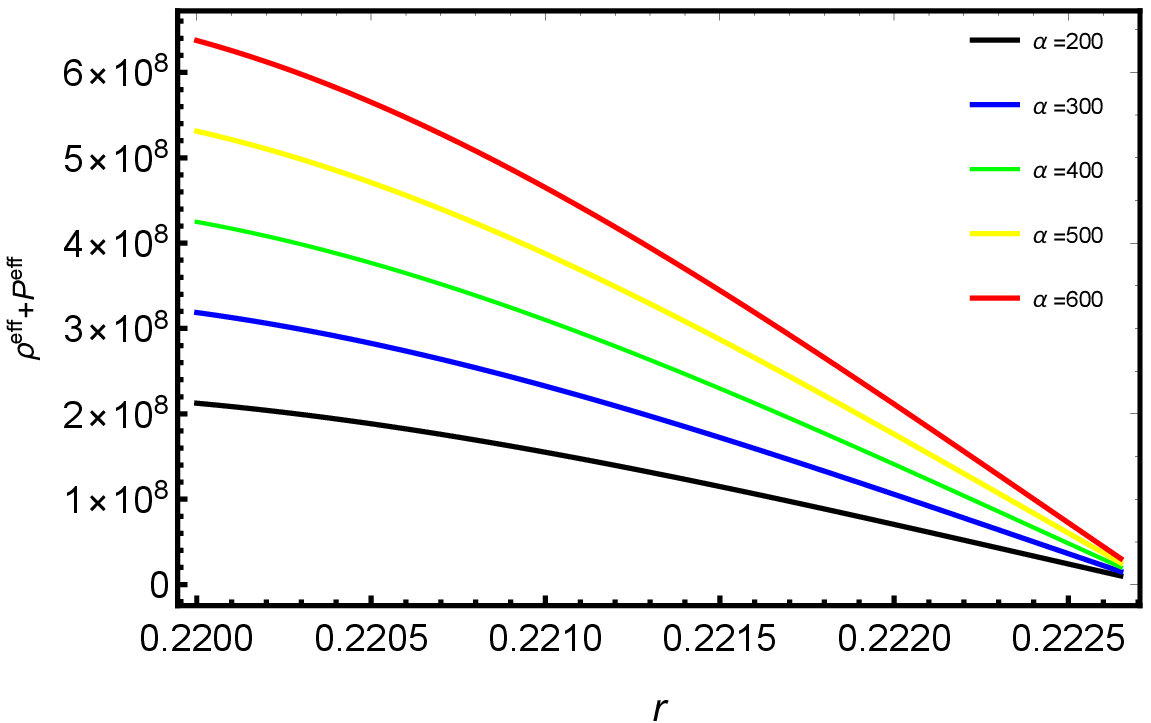,width=0.43\linewidth} &
\end{tabular}
\end{figure}
\FloatBarrier
\FloatBarrier

\subsection{ The Modified TOV Equation for $f(\mathcal{G})$ Gravity}

\begin{figure}[h!]
\begin{tabular}{ccccc}
\epsfig{file=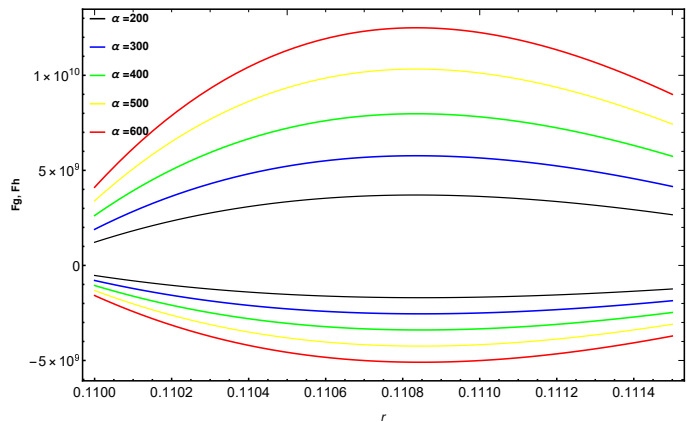,width=0.5\linewidth} &
\end{tabular}
\caption{Behavior of hydrostatic force $(\mathcal{F}_{h})$ and gravitational force $(\mathcal{F}_{g})$  under viable $f(\mathcal{G})$ gravity model.}
\label{Fig:4}
\end{figure}

The conservation equation of motion for our stellar system is given by
\begin{equation}\label{5.25}
\bigtriangledown^\mu T^{eff}_{\mu\nu}=0.
\end{equation}
Here, in this section we study the equilibrium condition of the stellar structure model. For this purpose we consider the modified form of TOV equation for an isotropic matter distribution, defined as
\begin{equation}\label{5.26}
-\frac{dp^{eff}}{dr}-\frac{\nu~'}{2}(\rho^{eff} +p^{eff})=0.
\end{equation}
For the physical viability of model, the equilibrium condition of compact configuration was proposed by
Tolman \cite{Tolman}, later on Oppenheimer and Volkoff \cite{Oppen} in such a
way that sum of the forces viz., gravitational force
$(\mathcal{F}_{g})$, hydrostatic force $(\mathcal{F}_{h})$ should be zero.
\begin{equation}\label{5.27}
\mathcal{F}_{g} +\mathcal{F}_{h}=0,
\end{equation}
where $\mathcal{F}_{g}= -\frac{\nu~'}{2}(\rho^{eff} +p^{eff})$, $\mathcal{F}_{h}=- \frac{dp^{eff}}{dr}$.
Here we also obtain fluctuating behavior if we plot for the full range of radius. This issue is due to the only conserved quantity obtained in our case. Further, it is observed that these
forces balance out each other effect for the smaller
range of radii, which justify the condition of equilibrium for our stellar system.
Hence, the behavior of forces $(F_g$) and $(F_h)$ is presented in FIG. $\ref{Fig:4}$, with one such small range.
The graphical representation of these forces for different values of model parameter is shown in FIG. $\ref{Fig:4}$, which clearly depicts that our system is in static equilibrium under the given forces.
\FloatBarrier

\subsection{Stability Analysis}

\begin{figure}[h!]
\begin{tabular}{cccc}
\epsfig{file=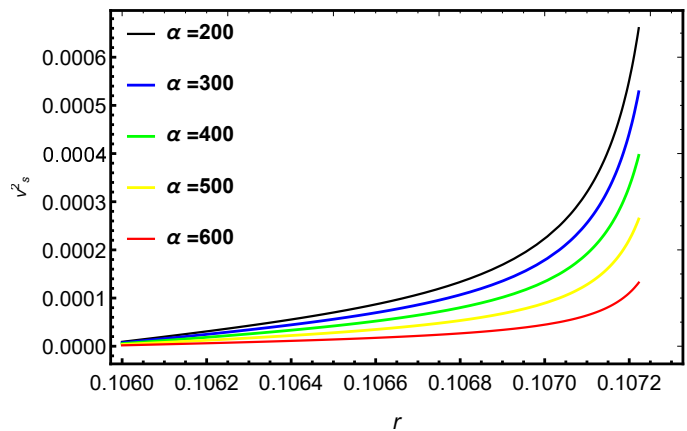,width=0.5\linewidth} &
\end{tabular}
\caption{Variation of $v^2_{s}$  under viable $f(\mathcal{G})$ gravity model.}
\label{Fig:5}
\end{figure}

Moreover, to examine physical acceptable and consistent model the role of stability for stellar structure has great significance.
In recent years, stability analysis has been discussed by many researchers.
To probe the stability of our proposed model, we consider the Herrera's cracking concept \cite{Her}. According to cracking concept, the square of sound speed symbolized by $v^2_{s}$ should satisfy the following stability condition i.e. $0\leq v^2_{s}\leq1$.
The sound speed is defined by the following relation.
\begin{equation}\label{5.28}
v^2_{s}=\frac{dp^{eff}}{d\rho^{eff}}.
\end{equation}
To preserve causality condition, the sound speed
must lie in interval $[0, 1]$ everywhere inside the star for a physically
stable stellar system.
The variation of  sound speed for the small range of radii can be easily seen from the FIG. $\ref{Fig:5}$. Here, we observe that matter configuration relation is stable, as discussed.
\FloatBarrier

\section{Concluding Remarks}

$\emph{Noether symmetry approach}$ is considered to be a fundamental tool for solving dynamical equations.
Moreover, their existence yields feasible conditions so that one may choose physically and analytically acceptable universe
models according to the recent observations.
Lagrange multipliers perform pivotal role to reduce the Lagrangian into its canonical form which is quite helpful to reduce the dynamics of the system for finding the exact solutions.
The main aim of this paper is to investigate the physical characteristics of compact star by exploiting the so-called $\emph{Noether symmetry approach}$. For this purpose, we consider the spherically symmetric spacetime and the functional form of modified $f(\mathcal{G})$ gravity in the context of isotropic matter distribution. We have derived the equation of motion and also constructed the expression for
point-like Lagrangian in the frame of modified $f(\mathcal{G})$ gravity.
Further, Noether symmetry generators along with associated conserved quantities have been computed by
considering the power law model i.e. $f(\mathcal{G})=\alpha \mathcal{G}^n$ with specific choice $n=2$.
Here, we can notice that Noether symmetry of spherically symmetric spacetime in modified $f(\mathcal{G})$ gravity
provide a rather handy conserved quantity. The existence of conserved quantity play
a phenomenal role in describing the structure of compact star.
For our case, the obtained conserved quantity is a highly nonlinear differential equation and it does not support any information
to solve this equation analytically. To our best of knowledge, this is the first effort to use conservation
relation obtained from the classical Noether approach, by imposing some specific initial conditions to develop the
expression for the metric potentials in the context of $f(\mathcal{G})$ gravity to discuss compact stellar structures.
The distinct major findings of present study provide physical justification, which are interpreted by the following properties related to these isotropic compact stars as follows:
\\
\begin{itemize}
\item The variation of metric potentials
$e^{\nu}$ and $e^{\lambda}$ with respect to the radial coordinate $r$ is shown in FIG. $\ref{Fig:1}$. The behavior of both metric potentials
have minimum value at the center of star and then increase monotonically away from the center to surface. For a physical viability and stability of the suggested models, metric potentials should be positive, finite and free from the geometrical singularities.
FIG. $\ref{Fig:1}$ clearly depicts that our metric potentials are consistent and stable.

\item The effective energy density and pressure
inside the stellar structure of star should be maximum at the center. We could not obtain clear plots for the full range $0\leq r\leq 10$. In fact, we come to know that favorable results are obtained in small pieces of radii.
However, we observe that at center these features attain maximum value and then instantaneously decrease towards the surface boundary, which is a positive indication for physically viable results through Noether symmetries. Hence, for clarity we have provided only the plots for the small range of radii to show the smooth behavior of our stellar structures by using Noether scheme.
It is clear from the FIG. $\ref{Fig:2}$  that effective energy density and pressure
attain maximum value and show continuously decreasing behavior when moving towards the boundary of the star.

\item It can be seen from the FIG. $\ref{Fig:3}$  that all energy bounds for our proposed model is well satisfied which exhibit
the realistic matter content.

\item  To check that whether all forces namely, gravitational force
$(F_g$) and hydrostatic force $(F_h)$ are in equilibrium for our proposed model, we studied TOV equation in modified $f(\mathcal{G})$ gravity frame of reference. FIG. $\ref{Fig:4}$  yields that the forces are in equilibrium, which guarantee the
stability of our system.
\item The speed of sound for compact star is denoted by $v^2_{s}$. From the FIG. $\ref{Fig:5}$, it can be easily seen that our model is consistent with the causality condition.
 \\
\end{itemize}

Conclusively, we observe that the stellar structures in $f(\mathcal{G})$ gravity from $\emph{Noether symmetry approach}$ depend on the conserved quantity obtained and the model parameter $\alpha$. For the present analysis, we have chosen possible value of the model parameter $\alpha$.
However, one can try some other $f(\mathcal{G})$ gravity models with the same approach with a hope to get more conserved quantities which might give some better physical results. As a final comment, we have successfully developed a stable and physically acceptable isotropic model via $\emph{Noether symmetry approach}$ in the context of modified $f(\mathcal{G})$ gravity. Here, we observe that all physical features of compact stellar structure follow physically accepted patterns, but with some restricted range of radii.\\\\


\section*{References}

\begin{thebibliography}{70}


\bibitem{14} S. Capozziello, R. Ritis, A. A. Marino, Class. Quant. Grav. \textbf{14}, 3259 (1997).

\bibitem{15} S. Capozziello, G. Marmo, C. Rubano et al., Int. J. Mod. Phys. D \textbf{6}, 491 (1997).

\bibitem{16} U. Camci, Eur. Phys. J. C \textbf{74}, 3201 (2014).

\bibitem{17} U. Camci, JCAP \textbf{07}, 002 (2014).


\bibitem{18} K. Bamba, S. Capozziello, S. Nojiri, S. D. Odintsov, Astrophys.
Space Sci. \textbf{342}, 155 (2012).

\bibitem{19} S. Capozziello, M. Laurentis, S. D. Odintsov, Eur. Phys. J. C \textbf{72}, 1434 (2012).

\bibitem{20} I. Hussain, M. Jamil, F. M. Mahomed, Astrophys. Space Sci. \textbf{337}, 373 (2012).

\bibitem{Noether} E. Noether, Invariante Variationsprobleme Nachr. D. Knig. Gesellsch. D. Wiss. Zu Gttingen, Math phys. Klasse, \textbf{235} (1918).

\bibitem{Hanc} J. Hanc, S. Tuleja, M. Hancova, American J. Phys. \textbf{72}, 428 (2004).


\bibitem{Noj1} S. Nojiri, S. D. Odintsov, Phys. Lett. B \textbf{631}, 1 (2005).


\bibitem{Cog1} G. Cognola, E. Elizalde, S. Nojiri, S. D. Odintsov, S. Zerbini, Phys.
Rev. D \textbf{73}, 084007 (2006).


\bibitem{Cog2} G. Cognola, E. Elizalde, S. Nojiri, S. D. Odintsov, S. Zerbini, Phys.
             Rev. D \textbf{75}, 086002 (2007).

\bibitem{T.Chiba} T. Chiba, JCAP \textbf{03}, 008 (2005).


\bibitem{Santos} J. Santos, J. Alcaniz, M. Rebou¸cas, N. Pires, Phys. Rev. D \textbf{76}, 043519 (2007).

\bibitem{Sen} A. Sen, R. J. Scherrer, Phys. Lett. B \textbf{659}, 457 (2008).


\bibitem{Noj2008}  S. Nojiri, S. D. Odintsov, P. V. Tretyakov, Prog. Theor. Phys. Suppl. \textbf{172}, 81 (2008).

\bibitem{Bamba} K. Bamba, S. D. Odintsov, L. Sebastiani, S. Zerbini, Eur. Phys. J. C \textbf{67}, 295 (2010).

\bibitem{eqiuv} K. Uddin, J. E. Lidsey, R. Tavakol, Gen. Relav. Gravit. \textbf{41}, 2725 (2009).



\bibitem{Sharif1} M. Sharif, S. Waheed, Can. J. Phys. \textbf{88}, 833 (2010).

\bibitem{Kucu} Y. Kucukakca, U. Camci,I. Semiz, Gen. Relativ. Gravit.\textbf{ 44}, 1893 (2012).

\bibitem{Cap2007} S. Capozziello, A. Stabile, A. Troisi, Class. Quantum Grav. \textbf{24}, 2153 (2007).

\bibitem{Bahamonde} S. Bahamonde, K. Bamba, U. Camci, JCAP \textbf{02}, 016 (2019).

\bibitem{Mush1} M. F. Shamir, M. Ahmad, Eur. Phys. J. C \textbf{77}, 55 (2017).

\bibitem{Mush2} M. F. Shamir, M. Ahmad, Mod. Phys. Lett. A \textbf{32}, 1750086 (2017).

\bibitem{Sharif2} M. Sharif, H. I. Fatima, J. Exp. Theor. Phys. \textbf{122}, 104 (2016).

\bibitem{Laurentis} M. De Laurentis, Physics Letters B \textbf{780} 205 (2018).

\bibitem{b} S. Bahamonde, K. Dialektopoulos, U. Camci, Symmetry \textbf{12}, 68 (2020).

\bibitem{s} A. N. Nurbaki, S. Capozziello, C. Deliduman, Eur. Phys. J. C \textbf{80}, 108 (2020).

\bibitem{f} M. F. Shamir, Eur. Phys. J. C \textbf{80}, 115 (2020).



\bibitem{Abbas1} G. Abbas, S. Nazeer, M. A. Meraj,  Astrophys. Space Sci. \textbf{354}, 449 (2014).

\bibitem{Abbas3} G. Abbas, A. Kanwal, M. Zubair, Astrophys. Space Sci. \textbf{357}, 109 (2015).

\bibitem{Cam} M. Camenzind,  Compact Objects in Astrophysics. Springer, Berlin (2007).

\bibitem{Abbas4} G. Abbas, D. Momeni, M. Aamir Ali, R. Myrzakulov, S. Qaisar, Astrophys. Space Sci. \textbf{ 357}, 158 (2015).

\bibitem{Ast2} A. V. Astashenok, S. Capozziello, S. D. Odintsov, J. Cosmol. Astropart. Phys. \textbf{01}, 001 (2015).

\bibitem{Ast3} A. V. Astashenok, S. Capozziello, S. D. Odintsov, Phys. Lett. B \textbf{742}, 160 (2015).

\bibitem{ Momen1} D. Momeni, P. H. R. S. Moraes, H. Gholizade, R. Myrzakulov, Int. J. Geom. Meth. Mod. Phys. \textbf{15}, 1850091 (2018).

\bibitem{Momen2} D. Momeni, H. Gholizade, M. Raza, R. Myrzakulov, Int. J. Mod. Phys. A \textbf{30}, 1550093 (2015).

\bibitem{Capo} S. Capozziello, M. D. Laurentis, R. Farinelli, S. D. Odintsov, Phys.
Rev. D \textbf{93} 023501 (2016).

\bibitem{Ast4} A. V. Astashenok, S. Capozziello, S. D. Odintsov, Phys. Rev. D \textbf{89},
103509 (2014).



\bibitem{Krori} K. D. Krori, J. Barua, J. Phys. A. Math. Gen. \textbf{8}, 508 (1975).

\bibitem{M.V} M. V. D. S. Silva, M. E. Rodrigues, Eur. Phys. J. C \textbf{78}, 638 (2018).

\bibitem{P.J} P. J. Olver, Applications of Lie Groups to Differential Equations (Springer Science  Business Media, New York, 2000).



\bibitem{Cap2010} S. Capozziello, M. De Laurentis, A. Stabile, Class. Quantum Gravity \textbf{27} (2010) 165008.

\bibitem{Tolman} R. C. Tolman, Phys. Rev. \textbf{55}, 364 (1939).

\bibitem{Oppen} J. R. Oppenheimer, G. M. Volkoff, Phys. Rev., \textbf{55}, 374 (1939).

\bibitem{Her} L. Herrera, Phys. Lett. A \textbf{165}, 206 (1992).


\end{thebibliography}
\end{document}